\documentclass[english]{article}
\usepackage[utf8]{inputenc}
\usepackage[T1]{fontenc}
\usepackage{babel}
\usepackage{amsmath}
\usepackage{graphicx}
\usepackage{url}
\usepackage[super,compress, comma]{natbib}
\usepackage{color, soul}
\usepackage{array}
\usepackage{fancyhdr}
\usepackage[notablist,nofiglist, nomarkers]{endfloat}

\newcolumntype{M}[1]{>{\centering\arraybackslash}m{#1}}
\begin{document}

\title{Access to mass rapid transit in OECD urban areas}

\author{Vincent Verbavatz\textsuperscript{1,2}, Marc Barthelemy\textsuperscript{1,3{*}}}

\maketitle
\thispagestyle{fancy}
~\\
1. Universit\'e Paris-Saclay, CNRS, CEA, Institut de physique th\'eorique, 91191, Gif-sur-Yvette, France. \\ 
2. Ecole des Ponts ParisTech, Champs-sur-Marne, France. \\ 
3. Centre d'Etude et de Math\'ematique Sociales, CNRS/EHESS, 54 Boulevard Raspail, 75006 Paris, France. \\
{*}corresponding author:
Marc Barthelemy (marc.barthelemy@ipht.fr)

\begin{abstract}
As mitigating car traffic in cities has become paramount to abate climate change effects, fostering public transport in cities appears ever-more appealing. A key ingredient in that purpose is easy access to mass rapid transit (MRT) systems. So far, we have however few empirical estimates of the coverage of MRT in urban areas, computed as the share of people living in MRT catchment areas, say for instance within walking distance. In this work, we clarify a universal definition of such a metrics - People Near Transit (PNT) - and present measures of this quantity for 85 urban areas in OECD countries – the largest dataset of such a quantity so far. By suggesting a standardized protocol, we make our dataset sound and expandable to other countries and cities in the world, which grounds our work into solid basis for multiple reuses in transport, environmental or economic studies.
\end{abstract}

\section*{Background \& Summary}

Motorized transport currently accounts for more than 15\% of world greenhouse gas emissions~\cite{herzog}. As most humans live in urban areas and two-thirds of world population will live in cities by 2050~\cite{un}, mitigating car traffic in cities has become crucial for limiting climate change effects~\cite{dodman, glaeser, Oliveira, newman1}. Daily mitigating is the main driver for passenger car use - about 75\% of American commuters drive daily~\cite{usdt} - while alternative transport modes such as public transportation networks are unevenly developed among countries and cities~\cite{wiki}. \\

Over the last decades, various attempts to assess the environmental impact of car use in cities have emerged from multiple fields, ranging from econometric studies to physics or urban studies~\cite{duranton, creutzig, pumain, barthelemy, newman2}. A seminal result of transport theory, by Newman and Kenworthy~\cite{newman2}, correlated transport-related emissions with a determinant spatial criterion: urban density. Alternatively, Duranton and Turner~\cite{duranton} claimed that public transport services were to unsuccessful in reducing traffic, as transit riders lured off the roads are replaced by new drivers on the released roads. Such results, however, crucially lack both theoretical and empirical foundations~\cite{buchanan, litman, anderson, baumsnow} and new research~\cite{verbavatz} shows that the two main critical factors that control car traffic in cities are urban sprawl and access to mass rapid transit (MRT). \\

More generally, understanding mobility in urban areas is fundamental, not only for transport planning, but also for understanding many processes in cities, such as congestion problems, or epidemic spread~\cite{dalziel,balcan} for example. But what is a good measure of access to transit? Studies have mainly focused on the number of lines or stops~\cite{fouracre,gallotti1,gallotti2}, length of the network or graph analysis~\cite{musso, gattuso, derrible}. Few works~\cite{idtp, singer, verbavatz}, however, have considered investigating catchment areas of MRT stations, i.e. looking at the share of population living close to MRT stations, for instance within walking distance. Such conditions have however proved to be essential in explaining commuting behaviours and mobility patterns~\cite{verbavatz}. \\

The most detailed definition of such catchment metrics is the People Near Transit (PNT), and originates from a 2016 publication from the Institute for Transportation and Development Policy (IDTP)~\cite{idtp}. It produces a rigorous dataset of the share of population living close to transit (less than 1~km) for 25 cities in the world (12 in OECD countries). However, definitions of urban areas and rapid transit systems in that dataset are multiple and need to be refined while the number of cities must be expanded. \\

Hence, in order to expand our global knowledge of urban mobility, we need a common, unified and universal definition of access to public transit as well as sound measures of such a quantity. In this paper, we clarify its definition and propose what is to our knowledge the largest dataset of PNT globally. \\

Our analysis uses functional urban areas (FUA) in OECD countries, a consistent definition of cities across several countries~\cite{djikstra}. We restrict our measures to mass rapid transit, usually referring to high-capacity heavy rail public transport, to which we added light rails and trams. In our sense, mass rapid transit thus encompasses:
\begin{itemize}
    \item Tram, streetcar or light rail services.
    \item Subway, Metro or any underground service.
    \item Suburban rail services.
\end{itemize}
Buses are not comprised in that definition. In contrast with~\cite{idtp}, we do not exclude any form of commuting trains based on station spacing or schedule criteria. As we detail it in the Method section, we identify services and corresponding stops with the General Transit Feed Specification (GTFS), a common format for public transportation schedules and associated geographic information~\cite{gtfs}. \\

\begin{figure}
    \centering
    \makebox[\textwidth][c]{\includegraphics[scale=0.6]{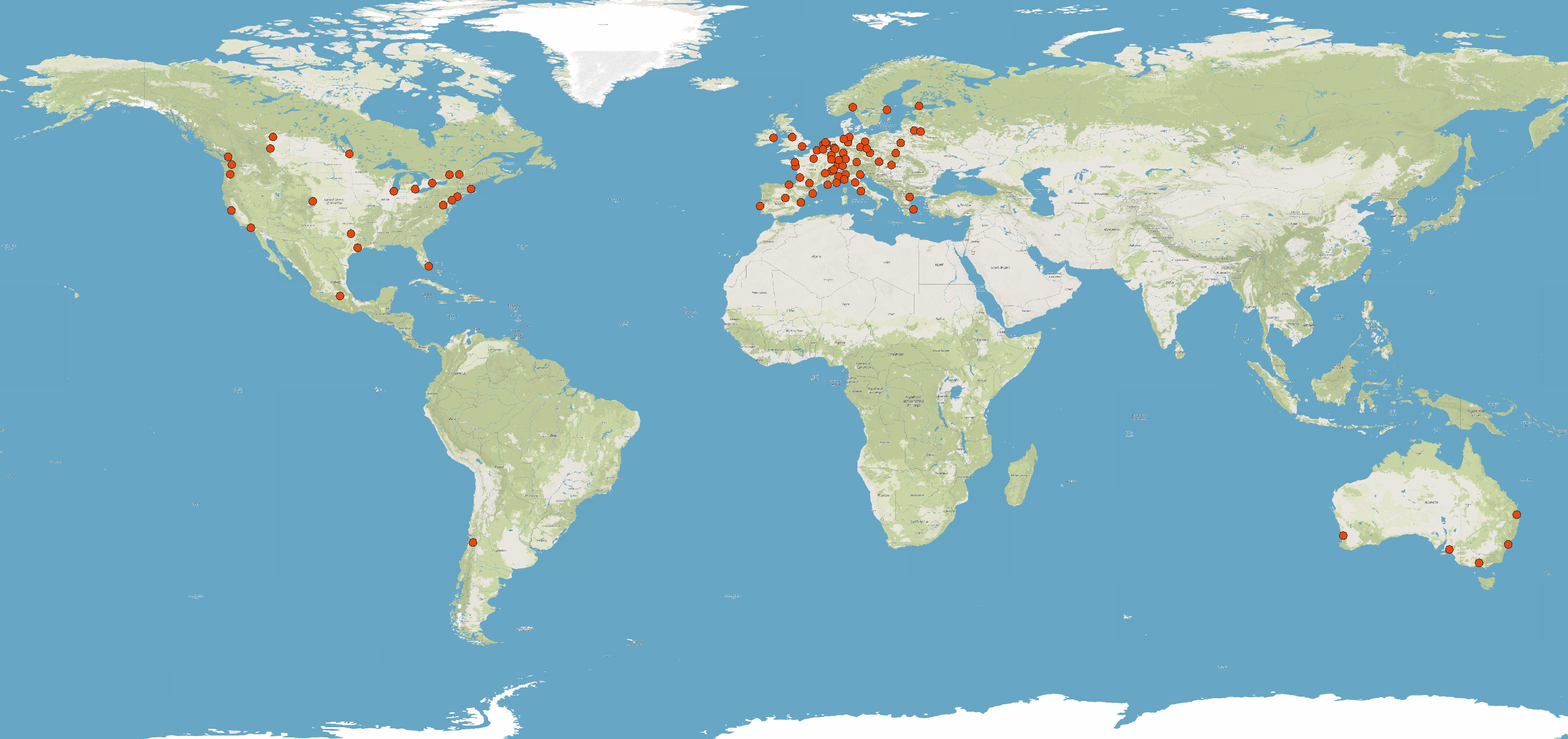}}
    \caption{The 85 OECD cities for which we found data are mostly found in Europe and in North America~\cite{maptiler}.}
    \label{map}
\end{figure}

Crossing open-access information from public transport agencies in OECD urban areas with population-grid estimates of world population~\cite{ghs}, we publish here a list of 85 OECD cities (see Fig. \ref{map}) for which we were able to compute the People Near Transit (PNT) levels defined as the share of urban population living at geometric distances of 500~m, 1,000~m and 1,500~m from any MRT station in the agglomeration:
\begin{align}
    \text{PNT}(d)=\frac{\text{population  s. t. } \text{cartesian minimum distance}<d}{\text{total population}}
\end{align}
where $d=500, 1000, 1500$. \\

We display on Tables~\ref{mrt-access-share-best} and \ref{mrt-access-share-poorest} the 5 cities with easiest access to MRT (largest PNT) and the 5 cities with scarcest access to MRT (smallest PNT). \\

\begin{figure}
    \centering
    \makebox[\textwidth][c]{\includegraphics[scale=0.6]{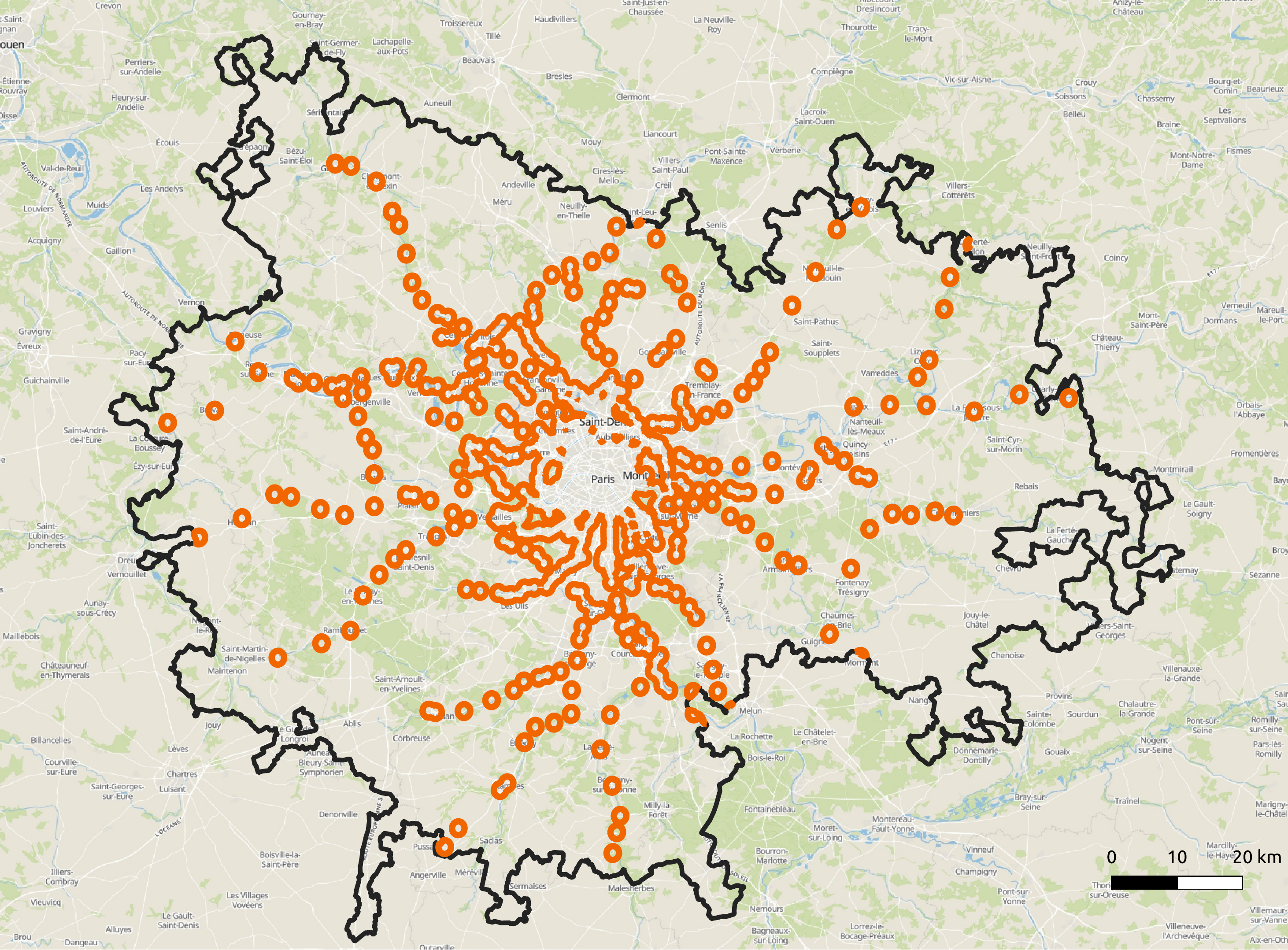}}
    \caption{1000~m catchment areas of MRT stations (in orange) in Paris functional urban area (boundaries are in black)~\cite{maptiler}.}
    \label{paris}
\end{figure}

We also provide for each city the population grid-maps with corresponding MRT access level, i.e. grid-maps of MRT catchment areas at different distances with population in each grid. As an example, Fig.~\ref{paris} shows the 1000~m catchment area of MRT stations in Paris. \\


\begin{table}
\centering
\begin{tabular}{|c|c|c|M{2cm}|M{2cm}|M{2cm}|}
\hline
\textbf{City}     & \textbf{Country} & \textbf{Population} & \textbf{500~m PNT (\%)} & \textbf{1000~m PNT (\%)} & \textbf{1500~m PNT (\%)} \\ \hline
Basel &	Switzerland	& 528811 &	57.78 &	80.15 &	86.96 \\ \hline
Bilbao &	Spain	& 986042 &	56.84 &	76.79 &	83.52 \\ \hline
Geneva	& Switzerland &	592893 &	50.44 &	74.68 &	85.07 \\ \hline
London	& United Kingdom &	11754700 &	43.09 &	72.56 &	85.8 \\ \hline
Zurich	& Switzerland & 	1329898 &	42.7 &	68.18 &	82.09 \\ \hline
\end{tabular}
\caption{Population Near Transit values: Share of population living within catchment area from a MRT station at thresholds 500~m, 1~000~m and 1~500~m. Top 5 cities with easiest (1000~m) access to MRT.}
\label{mrt-access-share-best}
\end{table}
\begin{table}
\centering
\begin{tabular}{|c|c|c|M{2cm}|M{2cm}|M{2cm}|}
\hline
\textbf{City}     & \textbf{Country} & \textbf{Population} & \textbf{500~m PNT (\%)} & \textbf{1000~m PNT (\%)} & \textbf{1500~m PNT (\%)} \\ \hline
Winnipeg &	Canada &	846133 &	0 &	0 &	0\\ \hline
Detroit	 &United States &	4263202 &	0.1 &	0.18 &	0.31\\ \hline
Houston	 &United States &	6706227	 &0.98 &	2.28 &	3.54\\ \hline
Miami	 &United Sates &	5964846	 &1.26 &	3.51 &	5.65\\ \hline
Dallas	 &United States &	7294931	 &1.18 &	4.05 &	7.64\\ \hline

\end{tabular}
\caption{Population Near Transit values: Share of population living within catchment area from a MRT station at thresholds 500~m, 1~000~m and 1~500~m. 5 cities with poorest (1000~m) access to MRT.}
\label{mrt-access-share-poorest}
\end{table}

\section*{Methods}

\subsection*{Residential populations for FUA}
Our analysis relies on the 2015 residential population estimates mapped into the global Human Settlement Population (GHS-POP) project~\cite{ghs}. This spatial raster dataset depicts the distribution of population expressed as the number of individuals per cell on a grid of cells 250~m long. Residential population estimates for the target year 2015 are provided by CIESIN GPWv4.10~\cite{GPW} and were disaggregated from census or administrative units to grid cells. \\

We downloaded population tiles that cover land on the globe in the Mollweide projection (EPSG:54009) and in raster format (.tif files). These raster data are made of pixels of width 250~m with associated value the number of people living in the cell. We processed the downloaded tiles with \textbf{Python} 3.7.6~\cite{python} and package \textbf{gdal} 3.0.2~\cite{gdal} to convert the raster files into vectorized shapefiles. The resulting shapefiles are comprised of polygons with field value the population in each polygon. Since the polygonization process merges adjacent pixels with common value into single polygons, populations for each polygon must be recomputed from polygon area and density through the simple following rule:
\begin{equation}
    \text{Pop}_{polygon} = \text{Pop}_{pixel} \times \frac{\text{Area}_{polygon}}{\text{Area}_{pixel}}
\end{equation}{}\\
where $\text{Area}_{pixel} = 250\times250=62~500$~m$^2$.
This leaves us with a list of 224 shapefiles of population that cover land area on earth. \\

By intersecting the resulting shapefiles with OECD shapefiles delineating Functional Urban Areas (FUA) in OECD countries~\cite{djikstra} (reprojected into Mollweide projection), we can build a population-grided dataset of cities in OECD countries. \\

These resulting files are the population substrates used for measuring population living close to MRT stations.

\subsection*{Extracts of MRT stations from GTFS files}
A common and \emph{de facto} standard format for public transportation schedules and associated geographic information is the General Transit Feed Specification (GTFS)~\cite{gtfs}. \\

A GTFS feed is a collection of at least six CSV files (with extension .txt) contained within a .zip file. It encompasses general information about transit agencies and routes in the network, schedule information such as trips and stop times and geographic information for stops (geographic coordinates). \\

The three main objects we require are:
\begin{itemize}
    \item Routes: distinct routes in the network of a certain type. A route is a (one-direction) regular line, for instance a metro or bus line. The route types we use are~\cite{gtfs}:
    \begin{itemize}
        \item Tram, Streetcar, Light rail. Any light rail or street level system within a metropolitan area.
        \item Subway, Metro. Any underground rail system within a metropolitan area.
        \item Rail. Used for intercity or long-distance travel.
        \item  Cable tram. Used for street-level rail cars where the cable runs beneath the vehicle, e.g., cable car in San Francisco.
    \end{itemize}
    Our definition of MRT excludes bus and ferry types:
    \begin{itemize}
        \item Bus. Used for short- and long-distance bus routes.
        \item Ferry. Used for short- and long-distance boat service.
    \end{itemize}
    \item Trips: trips are associated to a route and define a particular and scheduled trip between specific stations. For instance, the first train of the day is a trip.
    \item Stops: stops are geographic locations of the stops, stations and their amenities within the transit system. Stops are organized into a parent station and their amenities (e.g. platforms or exits).
\end{itemize}
Joining in this order the four tables \textbf{routes.txt},  \textbf{trips.txt},  \textbf{stop\_times.txt} and  \textbf{stops.txt} lets us bind stops with their associated route types. We can thus discriminate between bus stops and metro stops and thereby limit to our definition of MRT. \\

In a nutshell each GTFS file can be processed to produce localized and route-typed stops.

\subsection*{Measure of People Near Transit (PNT)}
In order to measure PNT within urban areas, we must bind transit systems with their respective FUAs. We need to retrieve - and merge - all available GTFS files pertaining to a specific urban area and make sure that no rapid transit agency is excluded in the process. \\

Most GTFS files for cities in the world are collected by the OpenMobilityData platform~\cite{OpenMobilityData}. For each city in our dataset, we cross-checked the OpenMobilityData with Wikipedia local network information \cite{wiki} to ensure that we considered all agencies of rapid transit within the urban area. \\

For some European countries (Germany, France), GTFS files were not availaible on OpenMobilityData and have been retrieved from other sources~\cite{de, fr}. We also note that GTFS format is not common in South Korea, Japan and in the United Kingdom where we only found GTFS data for Manchester area on OpenMobilityData~\cite{OpenMobilityData} while we directly used station coordinates for London \cite{tfl}. \\

We were thus left with a list of 85 urban areas in the world for which we had complete, reliable and extensive data. From route-typed stop coordinates within that dataset, we can extract MRT stops (excluding buses and ferries) and buffer - still using gdal - catchment areas for several distance thresholds: 500~m, 1~000~m and 1~500~m. Intersecting the resulting buffers with the population-grided shapefiles gives us the total population living within catchment areas, that can be expressed as a share of the total urban area population resulting in the value of the PNT metric. Our results are shown in Online-only Table~\ref{mrt-access-share}.

\subsection*{Code availability}
Detailed code generating the database can be accessed from the source code hosted via \emph{Gitlab}~\cite{gitlab}. \\

\section*{Data Records}
The Data Record of PNT in OECD urban areas is available online on \emph{Figshare}~\cite{datasource}. \\

PNT levels at distance thresholds: 500~m, 1~000~m and 1~500~m for the 85 Functional Urban Areas are shown on the Online-only Table~\ref{mrt-access-share}. The list of transit agencies for each city is  online along with PNT statistics (\textbf{mrt\_access.csv})~\cite{datasource}. \\

We also provide, for each city, grid-maps of population at different distances from MRT(\textbf{pops\_close\_to\_MRT.zip})~\cite{datasource}. \\

The Tables read as follows: Basel urban area has 528811 inhabitants, of which 57.78\% live within 500~m of a MRT station, 80.15\% within 1~000~m and 86.96\% within 1~500~m.

\begin{table}
\centering
\begin{tabular}{|c|c|c|M{2cm}|M{2cm}|M{2cm}|}
\hline
\textbf{City}     & \textbf{Country} & \textbf{Population} & \textbf{500~m PNT (\%)} & \textbf{1000~m PNT (\%)} & \textbf{1500~m PNT (\%)} \\ \hline
Adelaide          & Australia        & 1368481             & 11.43         & 27.9           & 40.79          \\ \hline
Amsterdam         & Netherlands      & 2766282             & 21.4          & 39.14          & 52.66          \\ \hline
Athens            & Greece           & 3667934             & 41.73         & 63.26          & 74.43          \\ \hline
Barcelona         & Spain            & 4838161             & 41.65         & 65.24          & 77.82          \\ \hline
Basel             & Switzerland      & 528811              & 57.78         & 80.15          & 86.96          \\ \hline
Berlin            & Germany          & 4953645             & 38.39         & 63.02          & 76.13          \\ \hline
Bilbao            & Spain            & 986042	& 56.84 &	76.79 &	83.52    \\ \hline
Bordeaux          & France           & 1176238             & 21.94         & 41.5           & 53.34          \\ \hline
Boston            & United States    & 4167892             & 13.4          & 30.69          & 44.69          \\ \hline
Bremen            & Germany          & 1253514             & 21.64         & 38.37          & 50.62          \\ \hline
Brisbane          & Australia        & 2307430             & 9.27          & 25.63          & 37.82          \\ \hline
Brussels          & Belgium          & 2632048             & 34.35         & 43.43          & 46.6           \\ \hline
Budapest          & Hungary          & 2972657             & 34.3          & 48.55          & 55.79          \\ \hline
Calgary           & Canada           & 1492971             & 5.7           & 18.93          & 31.15          \\ \hline
Chicago           & United States    & 9608320             & 5.91          & 13.34          & 18.45          \\ \hline
Cologne           & Germany          & 1960557             & 30.43         & 55.07          & 68.99          \\ \hline
Cracow            & Poland           & 1392519             & 22.36         & 34.41          & 41.95          \\ \hline
Dallas            & United States    & 7294931             & 1.18          & 4.05           & 7.64           \\ \hline
Denver            & United States    & 2738183             & 3.03          & 9.53           & 17.91          \\ \hline
Detroit           & United States    & 4263202             & 0.1           & 0.18           & 0.31           \\ \hline
Dresden           & Germany          & 1317454             & 35.47         & 56.14          & 66.95          \\ \hline
Dublin            & Ireland          & 1866112             & 15.96         & 35.23          & 50.17          \\ \hline
Dusseldorf        & Germany          & 1541332             & 33.01         & 55.86          & 69.01          \\ \hline
Edmonton          & Canada           & 1324949             & 3.6           & 10.8           & 17.26          \\ \hline
Florence          & Italy            & 770710              & 15.22         & 25.25          & 31.03          \\ \hline
Frankfurt am Main & Germany          & 2579579             & 28.35         & 56.11          & 72.33          \\ \hline
Geneva            & Switzerland      & 592893              & 50.44         & 74.68          & 85.07          \\ \hline
Genoa             & Italy            & 699462              & 11.3          & 26.02          & 36.9           \\ \hline
Hamburg           & Germany          & 3191585             & 16.6          & 39.5           & 55.71          \\ \hline
Hanover           & Germany          & 1272611             & 30.89         & 54.79          & 65.58          \\ \hline
Helsinki          & Finland          & 1451912             & 24.39         & 45.22          & 56.88          \\ \hline
Houston           & United States    & 6706227             & 0.98          & 2.28           & 3.54           \\ \hline
Kaunas            & Lithuania        & 380048              & 37.79         & 51.5           & 58.29          \\ \hline
Lausanne          & Switzerland      & 410089              & 32.63         & 65.0           & 80.17          \\ \hline
Leipzig           & Germany          & 972864              & 42.76         & 60.57          & 69.03          \\ \hline
Lille             & France           & 1360801             & 26.07         & 48.25          & 61.46          \\ \hline
Lisbon            & Portugal         & 2831367             & 18.17         & 41.66          & 55.96          \\ \hline
London       & United Kingdom    & 11754700 &	43.09 & 	72.56 &	85.80          \\ \hline
Los Angeles       & United States    & 17712325            & 2.26          & 8.56           & 15.91          \\ \hline
Luxembourg        & Luxembourg       & 577309              & 16.1          & 39.06          & 55.64          \\ \hline
Lyon              & France           & 1963944             & 29.56         & 50.85          & 63.87          \\ \hline
Madrid            & Spain            & 6615767             & 33.94         & 59.62          & 70.89          \\ \hline
\end{tabular}
\end{table}
\begin{table}
\renewcommand{\tablename}{Online-only Table}
\centering
\begin{tabular}{|c|c|c|M{2cm}|M{2cm}|M{2cm}|}
\hline
\textbf{City}     & \textbf{Country} & \textbf{Population} & \textbf{500~m PNT (\%)} & \textbf{1000~m PNT (\%)} & \textbf{1500~m PNT (\%)} \\ \hline
Manchester        & United Kingdom   & 3298781             & 16.34         & 43.57          & 64.61          \\ \hline
Marseille         & France           & 1779703             & 17.58         & 31.23          & 42.23          \\ \hline
Melbourne         & Australia        & 4466894             & 23.27         & 40.75          & 54.65          \\ \hline
Mexico City       & Mexico           & 20578866            & 8.8           & 20.6           & 28.89          \\ \hline
Miami             & United States   & 5964846             & 1.26          & 3.51           & 5.65           \\ \hline
Milan             & Italy            & 4966888             & 28.26         & 48.99          & 63.0           \\ \hline
Montreal          & Canada           & 4478991             & 10.88         & 25.82          & 37.75          \\ \hline
Munich            & Germany          & 2825789             & 36.74         & 62.44          & 75.31          \\ \hline
Nancy             & France           & 477056              & 16.42         & 36.16          & 52.43          \\ \hline
Nantes            & France           & 908423              & 21.41         & 40.08          & 50.82          \\ \hline
New York          & United States    & 19694439            & 27.02         & 45.35          & 56.68          \\ \hline
Nice              & France           & 848591              & 20.2          & 39.87          & 53.29          \\ \hline
Oslo              & Norway           & 1332133             & 30.16         & 47.59          & 53.67          \\ \hline
Ottawa            & Canada           & 1500455             & 2.18          & 7.46           & 12.03          \\ \hline
Paris             & France           & 12012223            & 37.16         & 62.7           & 77.56          \\ \hline
Perth             & Australia        & 1930198             & 4.88          & 15.66          & 27.01          \\ \hline
Philadelphia      & United States    & 6432106             & 4.55          & 14.34          & 23.81          \\ \hline
Portland          & United States    & 2262652             & 6.64          & 15.61          & 24.75          \\ \hline
Prague            & Czech Republic   & 2251032             & 35.21         & 59.46          & 73.52          \\ \hline
Rennes            & France           & 720142              & 10.18         & 23.61          & 35.69          \\ \hline
Rome              & Italy            & 4161006             & 25.93         & 46.97          & 58.33          \\ \hline
Rotterdam         & Netherlands      & 1823101             & 20.8          & 41.13          & 56.04          \\ \hline
San Francisco     & United States    & 6273368             & 10.92         & 25.15          & 37.74          \\ \hline
Santiago     & Chile &	7182609	& 13.0 &	34.01 &	49.49
  \\
 \hline
Seattle           & United States    & 3620117             & 2.83          & 6.45           & 10.12          \\ \hline
Stockholm         & Sweden           & 2221640             & 34.96         & 60.65          & 72.73          \\ \hline
Strasbourg        & France           & 779704              & 27.56         & 51.27          & 63.81          \\ \hline
Stuttgart         & Germany          & 2662983             & 27.57         & 51.06          & 64.64          \\ \hline
Sydney            & Australia        & 4903571             & 13.91         & 34.98          & 50.88          \\ \hline
Thessaloniki      & Greece           & 1076231             & 1.38          & 5.62           & 11.54          \\ \hline
Toronto           & Canada           & 7123826             & 13.21         & 23.06          & 33.72          \\ \hline
Toulouse          & France           & 1330243             & 14.27         & 30.02          & 42.65          \\ \hline
Turin             & Italy            & 1741546             & 38.38         & 52.55          & 61.3           \\ \hline
Utrecht           & Netherlands      & 882821              & 16.21         & 38.68          & 54.86          \\ \hline
Valencia          & Spain            & 1686890             & 31.04         & 56.66          & 71.73          \\ \hline
Vancouver         & Canada           & 2539976             & 8.52          & 22.55          & 34.37          \\ \hline
Venice            & Italy            & 557955              & 13.98         & 22.21          & 27.76          \\ \hline
Vienna            & Austria          & 2779253             & 44.63         & 68.02          & 78.94          \\ \hline
Vilnius           & Lithuania        & 691221              & 28.55         & 38.33          & 43.84          \\ \hline
Warsaw            & Poland           & 3099687             & 26.7          & 42.78          & 50.95          \\ \hline
Washington        & United States    & 8899517             & 3.12          & 8.42           & 12.89          \\ \hline
Winnipeg          & Canada           & 846133              & 0.0           & 0.0            & 0.0            \\ \hline
Zurich            & Switzerland      & 1329898             & 42.7          & 68.18          & 82.09          \\ \hline
\end{tabular}
\caption{Population Near Transit values: Share of population living within catchment area from a MRT station at thresholds 500~m, 1~000~m and 1~500~m for 85 OECD Functional Urban Areas.}
\label{mrt-access-share}
\end{table}

\section*{Technical Validation}
The most thorough and exhaustive measure of PNT in urban areas in existing literature is a 2016 report from the Institute for Transportation and Development Policy~\cite{idtp}. To validate our results and our methodology, we compared them with those results. \\

Out of the 12 OECD cities considered in~\cite{idtp}, 11 are in our dataset: 5 in the United States, 2 in Spain, 1 in Canada, 1 in France, 1 in the United Kingdom and 1 in the Netherlands (see Table~\ref{validation}). Unfortunately, we found no data in the remaining city: Seoul. \\

Out of these 11 cities, we had at first glance similar results for only two cities: Chicago (13\% for both) and Vancouver (19\% vs 23\%). The discrepancies observed for the other cases stem from different definitions of cities and from the different transit systems that were taken into account. While we work with Functional Urban Areas (FUA) only, the authors of \cite{idtp} mix two different definitions of cities: FUA and urban cores. By applying our method to urban cores and not functional urban areas, we found the same or similar results for Barcelona, Madrid, Rotterdam and Washington (see Table~\ref{validation}). \\

Also, the authors of \cite{idtp} considered a definition of the LRT (Light Rail Transit) and Suburban Rail that depends on station spacing and schedule criteria. We didn't choose this definition and for Boston and New York, we had therefore to exclude suburban trains - while keeping the definition of FUA - in order to retrieve  results similar to those of Table~\ref{validation}. In contrast, the study \cite{idtp} took into account the Bus Rapid Transit for Los Angeles, that we decided to exclude.  Finally, in Paris the authors of \cite{idtp} considered that the so-called RER trains were comprised in Suburban Rail, but not Transilien trains, while we included both systems in our analysis. \\

The conclusion here is that for similar definitions for cities and transit systems, we obtain similar results, validating our method and calculations. To facilitate comparison across future studies, we would recommend using the definition of cities given by Functional Urban Areas since it is very commonly used and already unified for OECD countries. Concerning transit systems, we think that it is more relevant and also verifiable to consider transit systems based on their types (Rail versus Road) rather that on spacing and schedule criteria that are specious and less universal. Hence, in comparing our results with results from the IDTP report~\cite{idtp} and after checking on Table~\ref{validation} that our methodology is correct, we decided to keep our unmodified estimations for the considered cities, despite the discrepancies with \cite{idtp}. \\

For other cities in the dataset we have unfortunately found no existing data to compare with. Thus, we hope for future research to test and expand our estimations and results.

\begin{table}
\centering
\begin{tabular}{|c|c|c|M{1.8cm}|M{1.0cm}|M{1.0cm}|M{1.0cm}|M{2.5cm}|}
\hline
  \textbf{City} & \textbf{Country} & \textbf{Population in}~\cite{idtp} & \textbf{Types in}~\cite{idtp}          & \textbf{PNT Share (\%) in}~\cite{idtp} & \textbf{Our PNT Share ($\%$)} & \textbf{Our PNT Share ($\%$) with}~\cite{idtp} \textbf{criteria} & \textbf{Comments about} \cite{idtp}                           \\
 \hline
  Barcelona     & Spain            & 3200000                  & Metro + LRT                  & 76                 & 65 & 74                 & Urban core and not FUA; suburban trains are \emph{de facto} included in~\cite{idtp}\\
 \hline
  Boston        & US               & 4650000                  & Metro + LRT                  & 15                 & 31 & 17                 & Excludes suburban trains                    \\
 \hline
  Chicago       & US               & 9500000                  & Metro                        & 14                 & 13 & 13                 &       /                                      \\
 \hline
  London   & UK             & 10000000                 & Metro + LRT + Suburban Rail            & 61                 & 73 & /                  & Some suburban trains are excluded in~\cite{idtp}  \\
 \hline
  Los Angeles   & US               & 13000000                 & Metro + LRT + BRT            & 11                 & 9 & /                  & We exclude Bus Rapid Transit                  \\
 \hline
 Madrid        & Spain            & 5500000                  & Metro + LRT                  & 76                 & 60 & 72                 & Urban core and not FUA; suburban trains are \emph{de facto} included in~\cite{idtp} \\
 \hline
  New York      & US               & 19800000                 & Metro + LRT                  & 35                 & 45 & 34                 & Excludes suburban trains                    \\
 \hline
  Paris         & France           & 12000000                 & Metro + Tram + Suburban Rail & 50                 & 63 & /                 & Some tramlines are excluded in~\cite{idtp} \\
 \hline
 Rotterdam     & Netherlands      & 1200000                  & Metro + LRT                  & 55                 & 41 & 50                 & Urban core and not FUA;  suburban trains are \emph{de facto} included in~\cite{idtp} \\
 \hline
  Vancouver     & Canada           & 2300000                  & Metro                        & 19                 & 23 & 23                 &       /                                      \\
 \hline
  Washington    & US               & 5800000                  & Metro                        & 12                 & 8 & 12                 & Urban core and not FUA                         \\
 \hline
\end{tabular}
\caption{Comparison of MRT Share from the IDTP report~\cite{idtp} with our estimations for 11 OECD cities. Discrepancies at first glance can be explained by different delineations of cities or transit systems. Applied on the same entities, results are similar.}
\label{validation}
\end{table}

\section*{Usage Notes}

Easy code and hints are given on \emph{Gitlab}~\cite{gitlab}. \\
~\\
We strongly recommand using GDAL \cite{gdal} to handle geographic data with Python.
\section*{Acknowledgements}
VV thanks the Ecole nationale des ponts et chauss\'ees for their financial support. This material is based upon work supported by the Complex Systems Institute of Paris Ile-de-France (ISC-PIF).

\section*{Author contributions}

VV and MB designed the study, VV acquired the data, VV analyzed and interpreted the data, VV and MB and wrote the manuscript.

\section*{Competing interests}

The authors have no competing interests.

\end{document}